# Reversible self-Kerr-nonlinearity in an N-type atomic system through a switching field

Xu Yang,[1] Kang Ying,[2] Yueping Niu,[1, *] and Shangqing Gong[1]

[1]*Department of Physics, East China University of Science and Technology, Shanghai 200237, China*
[2]*Shanghai Institute of Optics and Fine Mechanics,
Chinese Academy of Sciences, Shanghai 201800, China*

compiled: October 1, 2014

We investigate the self-Kerr nonlinearity of a four-level N-type atomic system in $^{87}$Rb and observe its reversible property with the unidirectional increase of the switching field. For the laser arrangement that the probe field interacts with the middle two states, the slope and the sign of the self-Kerr nonlinearity around the atomic resonance can not only be changed from negative to positive, but also can be changed to negative again with the unidirectional increasing of the switching field. Numerical simulation agrees very well with the experimental results and dressed state analysis is presented to explain the experimental results.

## 1. Introduction

Kerr-nonlinearity, which can be used in quantum nondemolition measurements [1, 2], quantum logic gates [3, 4] and the generation of optical solitons [5, 6], etc, has been a research hotspot in recent years. Many efforts have been made to achieve great enhancement of the Kerr-nonlinearity [7–12], among which the use of electromagnetically induced transparency (EIT) [13, 14] is a very important technology with the reduced absorption. In addition to the enhancement, the sign and slope modification of Kerr-nonlinearity also has specific applications in many fields, such as all-optical switch [15–17] and logic gates [18, 19] in optical communications and quantum computers. Recently, the group of Xiao changed the slope and sign of the self-Kerr-nonlinearity from negative to positive just by increasing the power of the additional switching laser [20]. In this work, we will report our investigation on the self-Kerr-nonlinearity in a similar N-type four level system but with the different laser fields arrangement. With the unidirectional increase of the switching laser power, the slope and sign of the Kerr-nonlinearity can not only be changed from negative to positive, but also can be changed to negative again. A dressed state analysis is given to demonstrate the above experimental observations.

## 2. Experimental arrangement

The N-type system used in our experiment is the D2 line (780nm) of $^{87}$Rb. The hyperfine states of $5^2S_{1/2}$ (F=1 and F=2) and $5^2P_{3/2}$ (F'=1 and F'=3) are used to form a four-level N-type configuration. Unlike the laser field arrangement in Ref. [20], here the coupling field $\Omega_c$ interacts with the ground state $|1\rangle$ and excited state $|3\rangle$ while the probe field $\Omega_p$ is applied between state $|2\rangle$ and $|3\rangle$. One may note that here the spontaneous emission from $|4\rangle$ to $|1\rangle$ is dipole forbidden and so will not show gain as [21] studied.

Figure 2 is our experimental setup, which is similar to the one used in our previous experiments [22–24]. The coupling, probe and switching fields are all single-mode tunable external cavity diode lasers (ECDL) (New Focus TLB-6900) with a linewidth of about 300 kHz. AOM is the Acousto-optic modulator, which can adjust the frequency of the probe field. The half-wave plate 1 (HWP1) and polarized beam splitter 1 (PBS1) are used to attenuate the power of the probe beam to below 50 $\mu$W to avoid the saturated absorption of the Rb atoms and self-focusing effect. About 10% of the coupling and switching laser power adjusted by HWP2, HWP3 and HWP4 are separated into an auxiliary Rb cell for stabilizing the frequency of the two beams using the saturated absorption spectroscopy (SAS) method. Then the three beams are brought together by PBS2 to interact with the Rb atoms. Before entering the cavity, the coupling and switching beams are focused by lenses with focal length of 30 cm to make their beam diameters be about 300 $\mu$m. PBS5 are used to reject the switching and coupling fields. The reflectivity of flat mirror M1 is approximately 99.5%. The cavity mirror M2 with a reflectivity of 99.5% is concave with a 15 cm radius of curvature and is controlled by a piezoelectric (PZT) driver. The finesse of the empty cavity (without the Rb cell and the PBS2) is about 138. After we insert the Rb cell and PBS2, the finesse of the cavity (with the Rb atoms off resonance) is reduced to 66 because of surface reflection losses.

---

* Corresponding author: niuyp@ecust.edu.cn

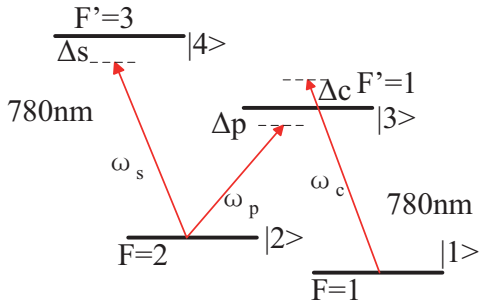

Fig. 1. N-type four-level atomic system in $^{87}$Rb and laser coupling scheme. $\Omega_p$, $\Omega_c$ and $\Omega_s$ refer to the switching, the probe and the coupling field. Here $\Delta_p = \omega_{32} - \omega_p$ is the probe detuning, and $\Delta_c = \omega_{31} - \omega_c$, $\Delta_s = \omega_{42} - \omega_s$ are the coupling and switching field detuning.

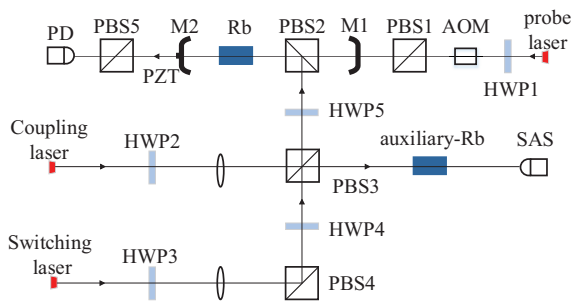

Fig. 2. Experimental setup. PD: photodetector, HWP: half-wave plate, PBS: polarizing beam splitter, PZT: piezoelectric ceramic transducer, M: mirror, AOM: acousto-optic modulator

In our experiment, first we lock the frequency of the coupling and switching fields to be resonant with the corresponding transitions using the SAS method. For the probe field, its frequency then can be tuned and locked by adjusting the AOM and SAS. Cavity scanning is realized by tuning the applied voltage of the PZT that attached to the cavity mirror M2. By scanning the cavity, the transmission spectrum shows a typical symmetric Lorentzian shape as the probe beam is tuned far from the atomic resonance. But when the probe field is tuned near the atomic resonance, the cavity transmission profile becomes asymmetric due to the self-Kerr-nonlinearity. Figure 3 shows the cavity transmission when $\Delta_p = 2$ MHz, $\Delta_c = \Delta_s = 0$, Pc=0.9 mW, Ps=0.7 mW. According to the relation between the nonlinear phase shift of the optical cavity and the nonlinearity of the inside medium, the self-Kerr nonlinear coefficient $n_2$ can be directly obtained from the asymmetry degree of the cavity transmission [25, 26]. Hence, we can measure the coefficient $n_2$ by scanning the cavity length through the PZT for each determined probe frequency. The probe field is totally tuned in the range of -50 MHz to 50 MHz. For different probe frequency, the intracavity power will change due to different absorption, so we should adjust the input probe laser power sightly in our experiment.

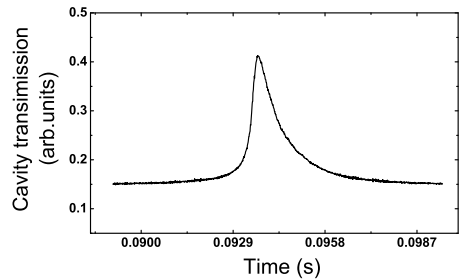

Fig. 3. Cavity transmission with the scan of the cavity length scanning. $\Delta_p = 2$ MHz, $\Delta_c = \Delta_s = 0$, Pc=0.9 mW, Ps=0.7 mW.

## 3. Comparison and Discussions

The left column of Fig. 4 shows our experimentally measured self-Kerr nonlinear coefficient $n_2$ at different probe detunings. From (a1) to (a5), the switching laser power Ps is increased from 0.3 mW to 2 mW while the coupling laser power Pc is kept at 0.9 mW. When Ps is much smaller than Pc, as (a1) and (a2) shows, the slope of $n_2$ near resonance is negative. But as Ps reaches 0.9 mW, the slope of $n_2$ near resonance changes from negative to positive (Fig. 4 (a3)). This phenomenon is similar to that of Ref. [20]. But what is worth pointing out is that as Ps increases further, the slope of $n_2$ near resonance goes back to be negative, which is different from that of Ref. [20]. As Fig. 4 (a4) and (a5) displayed, for cases of 1.2 mW and 1.5 mW, the slope of $n_2$ near resonance are both negative.

Based on the density-matrix equations that describe the motion of the atoms under the interaction of the three laser fields, we get the expression of self-Kerr nonlinearity through the steady-state solution under the weak-probe approximation [20, 27]. Also we have considered the Doppler effect and the collision decay in our numerical simulation. The right column of Fig. 4 represents our numerical theoretical result. One can find that the slope of $n_2$ near the resonance also changes from negative to positive and again to negative with the unidirectional increasing of the switching field power. This trend is in consistence with our experimental result. But one may note the magnitude of $n_2$ in (b3) is much larger than the other calculation results while the corresponding experimental magnitude is not so high. This can be explained by two reasons. First, from (b3) we can find the curve of $n_2$ has a very sharp change in a very small range of $\Delta_p$. However, the measurement step we chose is 3 MHz, so the place we measured is outside of the peak value. Second, according to our simulations, when $\Omega_s \neq \Omega_c$ or $\Omega_s$ is close to $\Omega_c$, the value of $n_2$ has a sharp decrease, as shown in the Fig. 5. In our experiment, it is hard to make sure that $\Omega_s = \Omega_c$ because the effective interaction ($\Omega_s/\Omega_c$) is determined by several parameters, including the light spot size, the dipole moment, etc. Therefore, the deviations of these factors may cause that $\Omega_s$ is not exactly equal to $\Omega_c$ in our experiment when Ps = Pc. As a result, there is difference between (a3) and (b3) in Fig. 4.

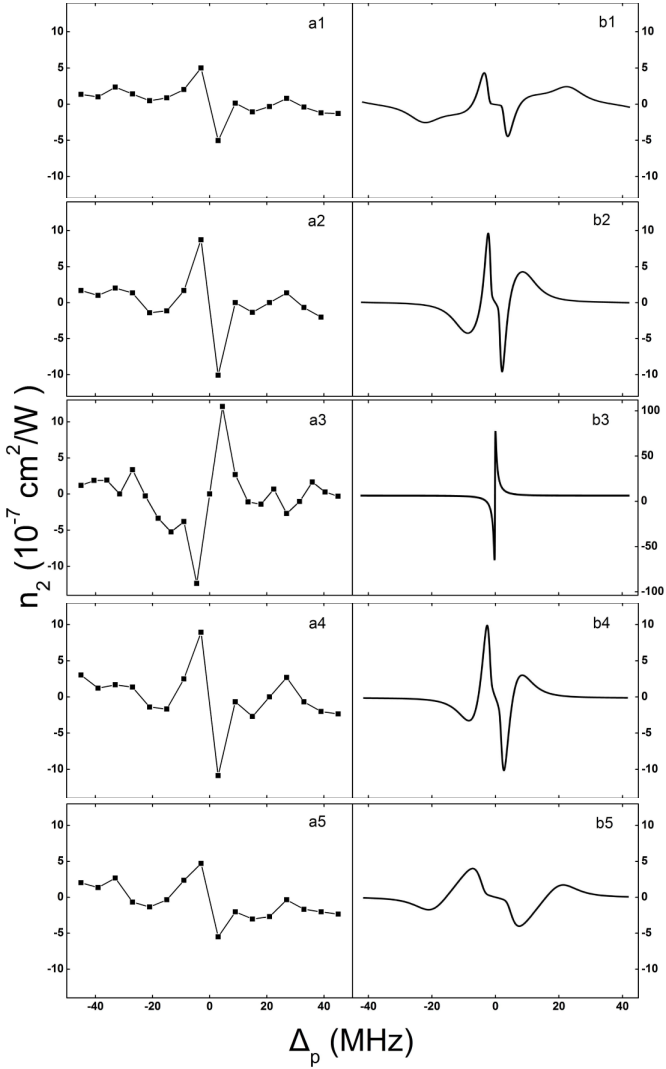

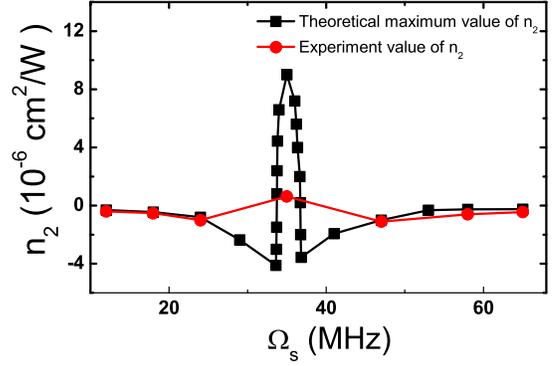

Fig. 5. Red spot is the experiment value of the self-Kerr nonlinear coefficient $n_2$ with different Ps we measured at about $\Delta_p = 2$ MHz, and black spot is the theoretical maximum value near the position of resonance with different $\Omega_s$. The other parameters used here are the same as Fig. 4.

Fig. 4. Left column: the experimental measured self-Kerr nonlinear coefficent $n_2$. Pc=0.9 mW, From (a1) to (a5), Ps=0.3 mW, 0.7 mW, 0.9 mW, 1.2 mW, 1.5 mW; Right column: the calculated self-Kerr nonlinear coefficient $n_2$ with parameters of $\Omega_c = 2\pi \times 35$ MHz and $\Omega_s = 2\pi \times 12$ MHz, $\Omega_s = 2\pi \times 24$ MHz, $\Omega_s = 2\pi \times 35$ MHz, $\Omega_s = 2\pi \times 47$ MHz, $\Omega_s = 2\pi \times 58$ MHz, $\gamma_{31} = \gamma_{32} = \gamma_{42} = 2\pi \times 3$ MHz, $\gamma_{41} = 0$. The collision decay is $\gamma^{col} = 2\pi \times 6$ MHz. T = 28°C.

When it comes to the sign of $n_2$, we take the value at about $\Delta_p = 2$ MHz as an example. Just as Fig. 5 shows, the red spot is the experiment value of the self-Kerr nonlinear coefficient $n_2$ and the black spot is the theoretical maximum value near the position of resonance with different $\Omega_s$. We can find $n_2$ changes from negative to positive and then reversed to negative with the unidirectional increasing of the switching laser power. But we did not experimentally get the maximum and minimum value of $n_2$. The reasons are just the same as the difference between (a3) and (b3) in Fig. 4, which have been mentioned above. However, we can still clearly see the reversible change of $n_2$ with the unidirectional increase of $\Omega_s$. Such change of sign or slope of nonlinear coefficient can be used in combined optical switches to form more complicated optical system, and it can also help to rich the modification methods of group velocity and optical solitons.

### 4. Dressed State Analysis

In this section, we will present the dressed state of this N-type system. For a weak probe field, levels $|1\rangle$ and $|3\rangle$ are dressed by $\Omega_c$ and levels $|2\rangle$ and $|4\rangle$ are dressed by $\Omega_s$ separately. The corresponding dressed states can be represented with:

$$|\alpha_{c1}\rangle = \sqrt{2}(|3\rangle + |1\rangle)/2$$
$$|\alpha_{c2}\rangle = \sqrt{2}(|3\rangle - |1\rangle)/2$$
$$|\alpha_{s1}\rangle = \sqrt{2}(|4\rangle + |2\rangle)/2$$
$$|\alpha_{s2}\rangle = \sqrt{2}(|4\rangle - |2\rangle)/2$$

The energy difference between the eigenvectors $|\alpha_{c1}\rangle \rightarrow |\alpha_{c2}\rangle$ and $|\alpha_{s1}\rangle \rightarrow |\alpha_{s2}\rangle$ are $\Omega_c$ and $\Omega_s$ respectively, just as Fig. 6(a) shows. Apparently, no matter $\Omega_s$ is larger or smaller than $\Omega_c$, there is always four absorption peaks. Thus, transparency occurs at the probe resonant position, where the slope of Kerr-nonlinear coefficient $n_2$ is negative. However, when $\Omega_s$ is equal to

$\Omega_c$, we can only see three absorption peaks. That is to say, the middle two absorption peaks just combine together as one, so that it turns to be absorptive at the probe resonant position where the slope of $n_2$ is positive. These agree very well to the observed experimental phenomenon in Fig. 4(a). Considering the natural linewidth and applied field, the absorption peak will have a certain width. So, when Ps is close to Pc the slope of $n_2$ near resonance is positive, and for the other Ps and Pc, it is negative.

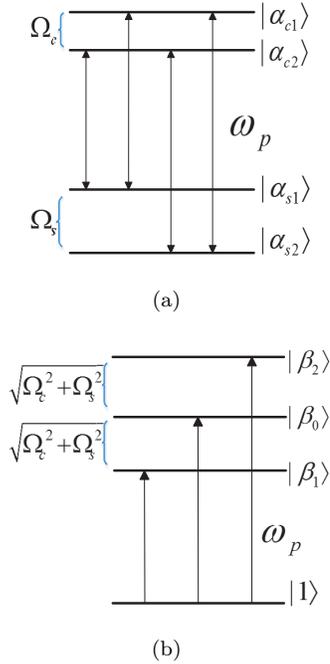

Fig. 6. (a) Dressed states for our N-type system as shown in Fig. 1; (b) Dressed states for the N-type system of Ref. [20].

If we use the above dressed state analysis to inspect the situation of the other laser arrangement of Ref. [20], we find that when $\Omega_s$ is small, it just slightly disturbs the EIT and the slope of $n_2$ near resonance is still negative. With the increasing of $\Omega_s$, it will dress the corresponding energy levels, along with $\Omega_c$. Then the three dressed states can be written as:

$$|\beta_0\rangle = (-\Omega_s|3\rangle + \Omega_c|4\rangle)/\sqrt{\Omega_c^2 + \Omega_s^2}$$

$$|\beta_1\rangle = (|2\rangle + (\Omega_c/\sqrt{\Omega_c^2 + \Omega_s^2})|3\rangle + (\Omega_s/\sqrt{\Omega_c^2 + \Omega_s^2})|4\rangle)/\sqrt{2}$$

$$|\beta_2\rangle = (-|2\rangle + (\Omega_c/\sqrt{\Omega_c^2 + \Omega_s^2})|3\rangle + (\Omega_s/\sqrt{\Omega_c^2 + \Omega_s^2})|4\rangle)/\sqrt{2}$$

The energy difference between the three eigenvectors are both $\sqrt{\Omega_c^2 + \Omega_s^2}$, as Fig. 6(b) shows. Hence it is absorptive at the probe resonant position. Therefore, the slope of $n_2$ will just change from negative to positive. Even though $\Omega_s$ is increased further, the slope of $n_2$ will keep to be positive.

## 5. Conclusion

In this paper, we have studied the self-Kerr nonlinear coefficient $n_2$ of the N-type system in detail. Since the probe field interacts with the middle two states $|2\rangle$ and $|3\rangle$, the slope and sign of $n_2$ near the resonance experience two dramatic changes between positive and negative with the unidirectional increase of the switching laser power, which shows the reversible property. Dressed state analysis has been presented to show clearly the experimental result.

This work was supported by the National Natural Science Foundation of China (Grant No. 11274112, 91321101) and the Fundamental Research Funds for the Central Universities (Grant No. WM1313003).